\documentstyle[12pt,fleqn]{article}
\begin{document}

\title{Effective widths and integrated cross sections
for two-photon transitions in $^{178}$Hf}

\author{Silviu Olariu and Agata Olariu\\
Institute of Physics and Nuclear Engineering,\\
Department of Fundamental Experimental Physics\\
76900 Magurele, P.O. Box MG-6, Bucharest, Romania\\
e-mail: olariu@roifa.ifa.ro}

\date{23 February 1999}

\maketitle

The possibility to produce an increase
of the activity of an isomeric sample by irradiating the sample with
a laser pulse and the x-ray pumping
of nuclear transitions have been discussed some time ago. \cite{1}
We have studied recently the rates for 
two-photon transitions from isomeric states to given final states,
induced by an incident flux of photons having a continuous energy distribution.
\cite{2},\cite{3}

In this work we have calculated integrated cross sections 
for two-photon processes in $^{178}$Hf. The calculations are based
on tabulated nuclear data. \cite{4}
We have assumed that the nucleus is initially 
in a state $|i\rangle$ of
energy $E_i$ and spin $J_i$, which in general is different from the
isomeric state.
By absorbing an incident photon  of energy
$E_{ni}$, the nucleus makes a transition to a higher
intermediate state $|n\rangle$ of energy $E_n$, spin $J_n$ and half-life $t_n$.
The state $|n\rangle$ then decays into a lower
state $|l\rangle$ of energy $E_l$ and spin $J_l$ 
by the emission of a gamma-ray photon having the  energy
$E_{nl}$ or by internal conversion.
In some cases the state $|l\rangle$ may be
situated above the initial state $|i\rangle$, and in these cases the
transition $|n\rangle\rightarrow |l\rangle$ is followed by further gamma-ray
transitions to lower states.

We have analyzed two-photon transitions in $^{178}$Hf for which there is an 
intermediate state $|n\rangle$ of known energy $E_n$, spin $J_n$ and 
half-life $t_n$, for which the 
intermediate state is connected by a known gamma-ray transition to the 
initial state $|i\rangle$ and to a lower state 
$|l\rangle$, and for 
which the relative intensities $R_{ni}, R_{nl}, R_{nl^\prime}$ 
of the transitions to lower states are known.

The cross sections are
summed over all lower states $|l\rangle$ which are possible 
for a given pair of initial and intermediate states $|i\rangle, |n\rangle$.
%of energies and spins 
%$E_i, J_i, 
%E_n, J_n,$ respectively. 
The integrated cross sections are calculated as 
%We calculated the total integrated cross section as
\begin{equation}
\sigma_{int}^{tot}=\frac{2J_n+1}{2J_i+1}\frac{\pi^2 c^2 \hbar^2}{E_{ni}^2}
\hbar \Gamma_{eff}^{tot} , 
\label{1}
\end{equation}
where
\begin{equation} 
\Gamma_{eff}^{tot}=\sum_{l \not= i} \Gamma_{eff} ,
\label{2}
\end{equation}
and where $E_{ni}=E_n-E_i$.
The quantity $\Gamma_{eff}$ is the effective width of the 
two-photon transition,
\begin{equation}
\Gamma_{eff}=F_R \ln 2/t_n ,
\label{3}
\end{equation} 
where $t_n$ is the half-life of the intermediate state $|n\rangle$, and
the dimensionless quantity $F_R$ has the expression
\begin{equation}
F_R=\frac{(1+\alpha_{nl})R_{ni}R_{nl}}
{\left[(1+\alpha_{ni})R_{ni}+(1+\alpha_{nl})R_{nl}
+\sum_{l^\prime} (1+\alpha_{nl^\prime})R_{nl^\prime}\right]^2} .
\label{4}
\end{equation}
In Eq. (\ref{4}),  
$\alpha_{ni}, \alpha_{nl},  \alpha_{nl^\prime}$ are
the internal conversion coefficients
for the transitions $|n\rangle\rightarrow |i\rangle,
|n\rangle\rightarrow |l\rangle, |n\rangle\rightarrow |l^\prime \rangle, 
l^\prime\not=i,l,$.  

The effective widths $\Gamma_{eff}^{tot}$ and the integrated cross sections
$\sigma_{int}^{tot}$ are given in Table I.
The multipolarities of the transitions are given in columns $ni$ and $nl$.
The largest integrated cross section found among
the listed 24 two-photon processes in $^{178}$Hf has the value 
$1.65 \times 10^{-26}$ cm$^2$ keV.
If the initial state is the isomeric state
at $E_i$=2446 keV, the nuclear parameters required for the calculation 
of the two-photon integrated cross section are known at present 
only for the intermediate state at $E_n$=2573 keV.

\begin{table}[h]
\caption{Effective width $\Gamma_{eff}^{tot}$ and integrated cross section
$\sigma_{int}^{tot}$ for two-photon transitions in $^{178}$Hf.}
\vspace*{0.5cm}
\begin{center}
\begin{tabular}{|c|c|c|c|c|c|c|}
\hline
$E_i$ & $E_n$ &$ni$ & $nl$ &$\Gamma_{eff}^{tot}$& $\sigma_{int}^{tot}$\\
(keV)&(keV)& & &(eV)  & (cm$^2$ keV)\\
\hline
1434.2	&	1496.4	&	E2	&	E2,M1	&	
3.32$\times 10^{-6}$	&	1.65$\times 10^{-26}$	\\
0	&	1174.6	&	E2	&	E2	&	
1.80$\times 10^{-4}$	&	2.51$\times 10^{-27}$	\\
0	&	1496.4	&	E2	&	M1,E2	&	
9.97$\times 10^{-5}$	&	8.56$\times 10^{-28}$	\\
93.2	&	1174.6	&	E2	&	E2	&	
1.79$\times 10^{-4}$	&	5.89$\times 10^{-28}$	\\
93.2	&	1496.4	&	M1	&	E2	&	
1.26$\times 10^{-4}$	&	2.46$\times 10^{-28}$	\\
0	&	1276.7	&	E2	&	M1,E2	&	
8.26$\times 10^{-6}$	&	9.73$\times 10^{-29}$	\\
93.2	&	1276.7	&	M1	&	E2	&	
1.63$\times 10^{-5}$	&	4.47$\times 10^{-29}$	\\
1513.8	&	1636.7	&	E1	&	E1,M1,E2	&	
1.29$\times 10^{-7}$	&	4.00$\times 10^{-29}$	\\
1554.0	&	1636.7	&	E1	&	E1,M1,E2	&	
7.57$\times 10^{-8}$	&	3.60$\times 10^{-29}$	\\
306.6	&	1276.7	&	E2	&	M1,E2	&	
1.18$\times 10^{-5}$	&	2.67$\times 10^{-29}$	\\
306.6	&	1496.4	&	E2	&	M1,E2	&	
1.26$\times 10^{-5}$	&	1.90$\times 10^{-29}$	\\
306.6	&	1174.6	&	E2	&	E2	&	
6.06$\times 10^{-6}$	&	1.72$\times 10^{-29}$	\\
1538.8	&	1636.7	&	M1	&	E1,M1,E2	&	
1.69$\times 10^{-8}$	&	8.27$\times 10^{-30}$	\\
1409.4	&	1636.7	&	E2	&	E1,M1,E2	&	
7.05$\times 10^{-8}$	&	6.41$\times 10^{-30}$	\\
1433.6	&	1636.7	&	E2	&	E1,M1,E2	&	
3.31$\times 10^{-8}$	&	4.85$\times 10^{-30}$	\\
1512.6	&	1636.7	&	M1	&	E1,M1,E2	&	
7.12$\times 10^{-9}$	&	1.78$\times 10^{-30}$	\\
306.6	&	1636.7	&	E1	&	E1,M1,E2	&	
2.85$\times 10^{-7}$	&	7.56$\times 10^{-31}$	\\
1384.5	&	1554.0	&	E2	&	E2,M2	&	
2.36$\times 10^{-10}$	&	4.55$\times 10^{-32}$	\\
632.2	&	1554.0	&	E2	&	E2,M2	&	
1.41$\times 10^{-9}$	&	6.37$\times 10^{-33}$	\\
306.6	&	1554.0	&	E2	&	E2,M2	&	
1.25$\times 10^{-9}$	&	4.46$\times 10^{-33}$	\\
1147.4	&	1554.0	&	M2	&	E2	&	
1.35$\times 10^{-10}$	&	2.40$\times 10^{-33}$	\\
2433.3	&	2573.5	&	M1	&	E2,M2	&	
6.14$\times 10^{-13}$	&	1.29$\times 10^{-34}$	\\
2446.1	&	2573.5	&	M2	&	M1,E2	&	
9.86$\times 10^{-14}$	&	2.09$\times 10^{-35}$	\\
2136.5	&	2573.5	&	E2	&	M1,M2	&	
5.93$\times 10^{-13}$	&	1.38$\times 10^{-35}$	\\
\hline
\end{tabular}
\end{center}
\end{table}

\end{document}